\documentstyle[12pt,epsf]{article}

\newcommand{\bmat}{\left(\begin{array}}
\newcommand{\emat}{\end{array}\right)}
\def\NPB#1#2#3{Nucl. Phys. B{#1} (19#2) #3}
\def\PLB#1#2#3{Phys. Lett. B{#1} (19#2) #3}

\def\PRD#1#2#3{Phys. Rev. D{#1} (19#2) #3}

\def\yzero{\smash{\hbox{$y\kern-4pt\raise1pt\hbox{${}^\circ$}$}}}

\def\beq{\begin{equation}}
\def\eeq{\end{equation}}
\def\beqa{\begin{eqnarray}}
\def\eeqa{\end{eqnarray}}

\def\-{\hphantom{-}}

\def\s2{\frac{1}{\sqrt2}}

\def\beq{\begin{equation}}
\def\eeq{\end{equation}}
\def\beqa{\begin{eqnarray}}
\def\eeqa{\end{eqnarray}}

\def\IF{\relax{\rm I\kern-.18em F}}
\def\II{\relax{\rm I\kern-.18em I}}
\def\IP{\relax{\rm I\kern-.18em P}}
\def\IC{\relax\hbox{\kern.25em$\inbar\kern-.3em{\rm C}$}}
\def\IR{\relax{\rm I\kern-.18em R}}

\def\Dsl{\,\raise.15ex\hbox{/}\mkern-13.5mu D} 
\def\IZ{Z\kern-.4em  Z}

\catcode`\@=11
\newdimen\@rotdimen
\newbox\@rotbox

\def\@vspec#1{\special{ps:#1}}
\def\@rotstart#1{\@vspec{gsave currentpoint currentpoint translate
   #1 neg exch neg exch translate}}
\def\@rotfinish{\@vspec{currentpoint grestore moveto}}
\def\@rotr#1{\@rotdimen=\ht#1\advance\@rotdimen by\dp#1%
   \hbox to\@rotdimen{\hskip\ht#1\vbox to\wd#1{\@rotstart{90 rotate}%
   \box#1\vss}\hss}\@rotfinish}
\def\@rotl#1{\@rotdimen=\ht#1\advance\@rotdimen by\dp#1%
   \hbox to\@rotdimen{\vbox to\wd#1{\vskip\wd#1\@rotstart{270 rotate}%
   \box#1\vss}\hss}\@rotfinish}%
\def\@rotu#1{\@rotdimen=\ht#1\advance\@rotdimen by\dp#1%
   \hbox to\wd#1{\hskip\wd#1\vbox to\@rotdimen{\vskip\@rotdimen
   \@rotstart{-1 dup scale}\box#1\vss}\hss}\@rotfinish}%
\def\@rotf#1{\hbox to\wd#1{\hskip\wd#1\@rotstart{-1 1 scale}%
   \box#1\hss}\@rotfinish}%
\def\rotate{\@ifnextchar[{\@rotate}{\@rotate[l]}}
\def\@rotate[#1]#2{\setbox\@rotbox=\hbox{#2}\@nameuse{@rot#1}\@rotbox}

\catcode`\@=12

\topmargin
-1.5cm
\textwidth
15.5cm
\textheight
23.5cm
\oddsidemargin
0.7cm
\evensidemargin
1.2cm

\begin{document}

\makeatletter
\@addtoreset{equation}{section}
\makeatother
\renewcommand{\theequation}{\thesection.\arabic{equation}}
\pagestyle{empty}
\rightline{FTUAM-99/15; IFT-UAM/CSIC-99-18}
\vspace{0.5cm}
\begin{center}
\LARGE{
Mirage gauge coupling unification 
\footnote{\bf This paper needs  substantial revision.
See the footnote in page 6.}
 \\[10mm]}
\large{
L.~E.~Ib\'a\~nez
\\[2mm]}
\small{
 Departamento de F\'{\i}sica Te\'orica C-XI
and Instituto de F\'{\i}sica Te\'orica  C-XVI,\\[-0.3em]
Universidad Aut\'onoma de Madrid,
Cantoblanco, 28049 Madrid, Spain.
\\[9mm]}
\small{\bf Abstract} \\[7mm]
\end{center}

\begin{center}
\begin{minipage}[h]{14.0cm}

We use compact  $D=4$, $N=1$, Type IIB orientifolds as a testing
ground for recent ideas about  precocious gauge coupling unification
and a  low energy string scale. We find that  certain such
orientifolds  have the interesting property
that gauge couplings receive  moduli-dependent corrections which mimic
the effect of field theoretical logarithmic running.
The effective cut-off  scale for the
logarithmic correction is not $M_{string}$ but
rather $M_X= \sqrt{\alpha}M_{Planck}M_c/M_{string}$,
where  $M_c$ is the compactification scale. Thus there is
just normal logarithmic running up to $M_{string}$ and
extra moduli dependent corrections which behave as if there was
further running to a higher virtual scale $M_X$.
In this mechanism a prominent role is played  by  anomalous
$U(1)$'s with moduli dependent Fayet-Iliopoulos terms.
A vanishing FI-term fixes the modulus dependence of
the corrected gauge coupling. We discuss possible ways to
implement this mechanism in the context of a simple
extension of the MSSM. Agreement identical (the
one-loop  equations are the same) to the one
obtained in SUSY-GUT's  is   obtained
 for $M_{string}=10^{11}$ GeV and
$M_c=10^{9}$ Gev. This fits  with the recent
suggestion, based on completely independent arguments,
 of identifying the string scale with the
intermediate scale.  Scenarios with
a 1 TeV string scale tend to yield too
small a value for $M_X$ in this context.

\end{minipage}
\end{center}
\newpage
\setcounter{page}{1}
\pagestyle{plain}
\renewcommand{\thefootnote}{\arabic{footnote}}
\setcounter{footnote}{0}

\section{Introduction}

There has been recently a lot of interest
in studying the viability of a  lowering of the scale of string theory.
Although the scale of string theory in perturbative heterotic vacua
is essentially fixed to be of order $gM_{Planck}$,  it has been
realized that it can be arbitrarily lowered in Type I and Type IIB
vacua \cite{witten}--\cite{ia2b} .
 This is possible because in the latter one can have the
charged matter fields living only on the worldvolume of some 3-branes
while gravity can propagate in all ten (or eleven ) dimensions.
Thus one can decouple the Planck mass from the string scale
by having some of the compact dimensions sufficiently large.

Although the possibility of lowering the string scale well below
the Planck mass is quite exciting, one of the least attractive
aspects of it is that the standard gauge coupling unification
of the MSSM is lost. Some ideas have been propossed to accomodate
precocious gauge coupling unification
\cite{bajogut,ghilen,BIQ,kakutev,cuiros,fram} but there is no longer a
prediction for gauge coupling unification which naturally fits
LEP data.

In order to study this issue further we believe it makes sense to
consider consistent explicit Type I $D=4$, $N=1$ vacua and
check the behaviour of gauge coupling unification in
specific models. The simplest such models are the
toroidal Type IIB, $D=4$, $N=1$ orientifolds \cite{orient} constructed in
refs.\cite{bl}--\cite{iru}  . They give rise to consistent four-dimensional
chiral theories with a variety of gauge groups.
A necesary  requirement to get chiral models is the location
of the relevant D-branes close to orbifold singularities.

It has been pointed out in refs.\cite{afiv,iru,imr}  that the gauge
kinetic functions corresponding to e.g., a set of D3-branes
at an orbifold singularity get a
piece proportional to the blowing-up fields $M$
of the given  singularity. In a simplified notation one
gets for the gauge kinetic functions a  general form
$f_b=S+c_bM$, where $S$ is the complex dilaton and
$c_b$ are constant coefficients. This structure is in fact
necesary in order to cancel $U(1)$ gauge anomalies \cite{iru}
present in these theories by a Green-Schwarz
mechanism \cite{gs} .
 Now, if the $c_b$ coefficients
were  proportional to the $\beta_b$-function coefficients
(and if $<ReM>\not=0$) ,
such structure could {\it mimic } some extra logarithmic running
and could modify substantially the conclussions about gauge
coupling unification. They behave like effective large
threshold corrections. It was argued in ref.\cite{imr}  that indeed,
in $Z_N$, $N$ odd compact orientifolds this is indeed the case
and the $c_b$ coefficients are proportional to the $\beta _b$'s.

In the present article we study that point in more detail.
The proportionality of those coefficients is due to the fact that
in such models the same $M$ fields which cancel some
anomalous $U(1)_a$ symmetries cancel at the same time certain
$\sigma $-model anomalies which are proportional to the
$\beta _b$ coefficients \cite{iru2} .  There is also a delicate interplay between
both type of anomalies. In particular, their simultaneous presence
makes  the $U(1)_a$ Fayet-Iliopoulos terms $\xi_a$ to get $T$(modulus)-dependent
corrections \cite{iru2} . More explicitly, one has a structure
of the type $\xi \propto (ReM-log(T+T^*))$. Upon minimization
one gets $<ReM>=log(T+T^*)$ and the gauge coupling gets
corrections $\propto \beta_b log(T+T^*)$. Upon considering
standard field theory running up to the string scale $M_{string}$
one finds that the {\it effective } large mass scale in the computation
is not $M_{string}$ but a scale $M_X=\sqrt{\alpha }M_{Planck}M_c/M_{string}$
where $M_c=1/R$ is the overall compactification scale.

Thus we claim that in $Z_N$ , $N$-odd compact orientifolds the couplings
at the string scale are {\it not} equal but are split in a manner which
actually {\it  mimics further running from $M_{string}$ up to a
virtual scale $M_X=\sqrt{\alpha }M_{Planck}M_c/M_{string}$}.
This is what we call "mirage" unification" : from low-energies
everything looks as if there was just standard field-theoretical
logarithmic running up to the scale $M_X$. In reality the
field theoretical running occurs only up to the string scale.
Unification is actually a mirage.

There is not at present any $D=4$, $N=1$, Type IIB orientifold
with a completely realistic spectrum. However
one can contemplate the possibility that a unification mecanism
like the one for the above  orientifolds could be at work for a
realistic model including the MSSM. In this case we find that
the experimentally preferred scale $M_X=10^{16}$ GeV is only obtained
for $M_{string}=10^{11}$ GeV and $M_c=10^{9}$ GeV.
Remarkably enough these are the values for the fundamental scales
recently propossed in ref.\cite{BIQ}  on the basis of completely different
arguments. Indeed, in that reference it was propossed the identification
of the string scale with the intermediate scale
$M_{string}=\sqrt{M_{Planck}M_W}=10^{11}$ GeV in order to understand
the generation of the $M_{Planck}/M_W$ hierarchy in terms of the
ratio $M_{string}/M_c$.

The structure of this article is as follows. In the next section
we describe briefly the cancelation of gauge $U(1)$ anomalies
and $\sigma$- model anomalies in $Z_N$ compact orientifolds with
odd $N$. In section 3 we describe the structure of the
Fayet-Iliopoulos terms asociated to the anomalous $U(1)$'s
and show how their cancellation leads to the
"mirage" unification described above.
In section 4 we describe the structure of a simple extension of
the MSSM including the mirage unification mechanism. In its simplest form
it will require the existence of an anomalous $U(1)$ whose
mixed anomalies with the groups of the standard model coincide
with the respective $\beta $-functions.
We give  in section 5 the final comments and conclussions
and discuss about the possible generality of these results.

\section{  $U(1)$'s and $\sigma$-model anomalies  in Type IIB, $D=4$,
$N=1$
orientifolds}

In our discussion a prominent role is played by both
anomalous $U(1)$'s and $\sigma $-model symmetries of the theory.
Type IIB $D=4$, $N=1$ orientifolds
  have generically $U(1)$ gauge
interactions whose triangle anomalies are non-vanishing \cite{iru}
\footnote{For a more phenomenological description of these
models see \cite{imr}  and references therein.} . These anomalies
are cancelled by a generalized Green-Schwarz mechanism which involves the
exchange of twisted singlet fields $M_f$ asociated to the fixed
points $f$ under the orbifold action \cite{iru}
. To be specific, let us consider
the case of $Z_N$ , $N$-odd orientifolds. Asociated to each fixed
point $f$ (e.g., 27 of them for $Z_3$ and 7 for $Z_7$) there are
twisted moduli fields $M_f^k$, with $k=1,..,(N-1)/2$.
The gauge kinetic function has a $M_f^k$-dependent piece
which appears at the disk level \cite{iru,iru2}
 \footnote{We are defining here
$ReS=8\pi ^2/g^2$.}
\beq
 f_b = S + {1\over N}     \sum_{f} \sum_{k=1}^{(N-1)/2}
{ {\cos(4\pi kV_b) }\over {C_k} }  M_f^k
\label{ffunc}
\eeq
where $C_k=\prod_{i=1}^3 2sin(\pi kv_i)$ and  $v_i$ are   the twist
eigenvalues
of the orbifold along the $i-th$ complex direction
\footnote{Thus $v=1/3(1,1,-2)$ for $Z_3$ and $v=1/7(1,2,-3)$ for $Z_7$.}.
One can check that  $C_k^2$ equals the number of
fixed points of the orbifold.
The $V_b$'s  correspond  to fractional numbers $l/N$ which are model
dependent.
Thus e.g., in the case of $Z_3$ in which the gauge group is
$U(12)\times SO(8)$ one has $V_{SU(12)}=1/3$ and $V_{SO(8)}=0$
(see ref.\cite{afiv} for examples and notation).
Now, under a $U(1)_a$ gauge transformation with parameter $\Lambda _a(x)$
the twisted $M_f^k$ fields transform nonlinearly in the fashion:
\beq
Im M_f^k \rightarrow  Im M_f^k\ + \ \delta _{GS\, k}^a  \
\Lambda_a(x)
\label{shift}
\eeq
with
\beq
\delta _{GS\, k}^a\ = \ 2n_a \sin(2\pi kV_a)
\label{deltags}
\eeq
Here $n_a$ is the rank
of the $U(n)$ or $SO(2n)$ group involved.
One can check that indeed this transformation of the $M_f^k$ fields
 combined with eq.(\ref{ffunc}) exactly cancels the mixed gauge anomalies
between the $U(1)_a$ field and the non-Abelian factors $G_b$.
Notice that in these models, unlike the heterotic orbifold models,
there may be more than one anomalous $U(1)_a$ whose anomaly
is cancelled by this mechanism \cite{afiv,fin, iru} .

These $Z_N$, odd $N$ orientifolds, like their heterotic counterparts,
have also certain $\sigma $-model invariances
\cite{modul} . Indeed, the Kahler potential
associated to the complex dilaton $S$, the three diagonal untwisted
moduli $T_i$ and the charged fields $A_i$ associated to the open strings
has the form:
\beq
K(S,S^*,T_i,T_i^*)\ =\ - log(S+S^*)-\sum_{i=1}^3\log(T_i+ T^*_i -|A_i|^2)
\label{kahlerunt}
\eeq
The effective classical action presents a $\sigma $-model invariance
under $SL(2,R)_{T_i}$ transformations given by
\footnote{These are not expected to be exact symmetries of the theory,
very much like in their heterotic duals,  where it is well known that
only the discrete subgroup $SL(2,{\bf Z})_{T_i}$ survives. Still,
as in the heterotic models \cite{modul} , cancelation of $\sigma $-model
anomalies is expected.}
\beq
T_i\rightarrow{a_iT_i-ib_i\over ic_iT_i+d_i},
\label{dual}
\eeq
with $a_i,b_i,c_i,d_i\in{\bf R}$ and $a_id_i-b_ic_i=1$.
Under these transformations the charged matter fields $A_j$
transform as:
\beq
A_j \rightarrow A_j  \prod_{i=1}^3(ic_iT_i+d_i)^{n_j^i}
\label{transc}
\eeq
where $n_j^i=-\delta_j^i $.
These transformations induce chiral rotations in the massless
fermions of the theory.
 They are asociated to gauge transformations of a composite
gauge vector potential involving the moduli fields $T_i$
\cite{modul} . If we
compute
the triangle anomalies corresponding to this composite current
and two gauge currents one finds in general an anomalous result.
The coefficient of this anomaly can be computed to be given
by \cite{modul} :
\beq
{b'}_a^i=-C(G_a)+\sum_{\underline R_a}
T(\underline R_a)(1+2n_{\underline R_a}^i)
\label{coeff}
\eeq
Here $C(G_a)$ is the quadratic Casimir of the gauge group
$G_a$ in the adjoint representation and $T(\underline R_a)$ is the
quadratic Casimir in the representation ${\underline R}_a$ corresponding
to a charged field.
It has been recently argued \cite{iru2} that these $\sigma $-model anomalies
may be cancelled again by a Green-Schwarz type of mechanism,
at least in the case of odd order $Z_N$ orientifolds
\footnote{See ref.\cite{iru2}  for a discussion of the $N$ even case.}
. Indeed, the $\sigma $-model-gauge mixed anomalies may be
cancelled if the twisted $M_f^k$ fields transform under
$SL(2,{\bf R})_{T_i}$ like :
\beq
Im M_f^k \rightarrow  Im M_f^k\ + \  \ \delta_{GS\, k}^i
 \log(ic_iT_i+d_i)
\label{shiftm}
\eeq
where
\beq
\delta_{GS\, k}^i\ =\  2tg(\pi k v_i)
\label{deltagsm}
\eeq
Indeed, as shown in ref.\cite{iru2}  , the anomaly coefficients
$b_a^i$' can be reexpressed as:
\beq
b_a^{i\, '} \ = \  -{{2}\over N}\ \ \sum_{k=1}^{(N-1)/2} \ C_k\
tg(\pi kv_i)  \ \cos4\pi kV_a
\label{masterorientm}
\eeq
which is exactly cancelled by the mechanism discussed above. Indeed, the
transformation (\ref{shiftm}) applied to eq.(\ref{ffunc}) gives
precisely a piece which cancels eq.(\ref{masterorientm}).

\section{Fayet-Iliopoulos terms and mirage unification}

It is well known that, whenever anomalous $U(1)$'s are present,
associated Fayet-Iliopoulos term in general appear
both in the heterotic case \cite{fi} and in
$D=4$, $N=1$  IIB orientifolds \cite{afiv,iru,iru2,pop,cvenil}  .
In a model with both anomalous $U(1)$'s and anomalous
$\sigma $-model symmetries
the invariance under the transformations in (\ref{shift}) and (\ref{shiftm})
requires that the Kahler potential of the $M_f^k$ fields has the general
invariant form \cite{iru2} :
\beq
K(M_f^k,{M_f^k}^*)\ =\ K(M_f^k+{M_f^k}^*
\ - \sum_a \delta_{GS \, k}^a  V_a  \ +
\  \sum_{i=1}^3 \delta_{GS\, k}^i log(T_i+T_i*) )
\label{kaememas}
\eeq
For a cuadratic Kahler potential \cite{pop} for the
$M_f^k$ fields, eq.(\ref{kaememas}) gives rise to a FI-term
corresponding to the $U(1)_a$ field :
\beq
\xi_a \ =\ -  \sum_f \sum_k (  \delta_{GS \, k}^a \ (M_f^k+ M_f^{k\, *} )
\ +  \ \sum_{i=1}^3 \delta_{GS\, k}^i log(T_i+T_i^*) )
\label{fiagranel}
\eeq
Notice that, unlike the case of non-compact orientifolds, here the
Fayet-Iliopoulos terms get a $T_i$-dependent piece \cite{iru2} .
This  piece is the one we want to discuss now.
Notice that, in the absence of non-Abelian gauge symmetry breaking
(which is what we want if we are interested in studying
 the corrections to the
couplings of the {\it initial} unbroken group), the
scalar potential will have minima at:
\beq
ReM_f^k\ =\  {-1\over 2}\sum_i \delta_{GS\, k}^i log(T_i+T_i^*) \ .
\label{maspunto}
\eeq
Plugging this expression into the real part of eq.(\ref{ffunc})
one gets:
\beq
 {{8\pi^2}\over g_a^2}\  = ReS -  {1\over {2N}}
\sum_i      \sum_{f} \sum_{k=1}^{(N-1)/2}
{ { \cos(4\pi kV_b)} \over {C_k}}
 \delta_{GS\, k}^i  log(T_i+T_i^*)
\label{ffuncdes}
\eeq
Thus we observe that, at SUSY-preserving vacua with $\xi_a =0$, there
are corrections to the gauge coupling constants which may be expressed in terms
of the untwisted moduli $T_i$ (or, rather, their vevs).
Let us now for simplicity consider the behaviour with respect to the overall
modulus field $T=T_i$ for all $i=1,2,3$. Then one can write:
\beq
 {{8\pi^2}\over g_a^2}\  = ReS \ - \ log(T+T^*){1\over {2N}}  \sum_i   \sum_{f}
\sum_{k=1}^{(N-1)/2}
\cos(4\pi kV_b) {{2tg(\pi k v_i)}\over{C_k} }
\label{ffuncdes2}
\eeq
Now, recalling eq.(\ref{masterorientm}) and the fact that
 $C_k^2$  equals  the number of fixed
points  one realizes
\footnote{This formula is incomplete since it only contains the
contribution to the coupling from the Wilsonian piece of the action.
When one includes the effect of the rescaling of the kinetic terms 
of massless fields, an extra piece given by 
${-1\over 2} \beta _a  log(T+T^*)$ has to be added, as pointed out 
recently in ref.\cite{abd} . This cancels 
the term coming from $<ReM>$.  Thus, contrary to the claim below,
there is actually no "mirage unification" in $Z_N$, $N$ odd
orientifolds, unification takes place at the string scale.
It is unclear what will happen in other orientifolds 
corresponding to D-branes sitting at singularities. In more
general cases one expects also a general structure 
$f_a=S+s_aM$ for the gauge kinetic function although now 
$s_a$ is not necesarily given by the beta-function as in
$Z_N$,  $N$ odd orientifolds. One also expects $M$ to mix
with $log(T+T^*)$ but a perfect cancellation between 
the ${-1\over 2} \beta _a  log(T+T^*)$ piece from the rescaling 
and the term from $<ReM>$ is in general not expected. 
Thus large corrections for the gauge coupling constants 
coming from this misscancelation will in general be present.
This may help in making compatible the existence of a low
string scale with the coupling unification problem.
In particular, if the $s_a$ coefficients were proportional
to $b_a$,these corrections could be reabsorved into
a redefinition of $M_X$ and the mirage unification 
scenario discussed in the text could  be realized.
However,
this possibility is not exemplified by $Z_N$ , $N$ odd 
orientifolds, as wrongly stated in this paper. A 
properly revised version will be submitted in due course.} 
:
\beq
 {{8\pi^2}\over g_a^2}\  = ReS \ + \ {1\over 2} \beta _a  log(T+T^*)
\label{ffuncdes3}
\eeq
where we have made us of the fact that $\sum_ib_a^i$'=$\beta _a$,
the corresponding $\beta $-function. This can be easily checked
from eq.(\ref{coeff}).
Eq.(\ref{ffuncdes3})  is an interesting result because it shows us that for
SUSY vacua with vanishing FI-terms the twisted-moduli-dependent
piece of the gauge coupling constant may be re-expressed in terms
of the untwisted moduli with  a coefficient that is no other but
the $\beta$-function.
It is this fact which leads to the "mirage unification" that we
mentioned above.

Indeed, let us now add the effect of the
field theory  running of couplings up to a
 cut-off equal
to the string scale:
\beq
{{8\pi ^2 }\over {g_a^2(Q^2)}} \ =\ ReS\  +\  {1\over 2} \beta_a
log(T+T^*)
\ +\ {1\over 2}  {{\beta _a}}log{{M_{string}^2}\over {Q ^2}}
\eeq
Now, setting as the cut-off for logarithmic running $M_{string}$
is only correct if there are no other thresholds of charged
particles at lower energies. Thus if we are working with 9-branes this would
require going to compactification scales above $M_{string}$. But in
that case we better do a T-duality transformation and work with 3-branes.
Then we would have the compactification scale $M_c$ below $M_{string}$ but
that would cause no new charged threshold.
 For 3-branes we know that  (see e.g. ref.\cite{imr} )
$T+T^*=2M_{string}^4/(M_c^4\alpha )$, with $\alpha =g^2/4\pi $,  and hence
we  have now:
\beq
{{8\pi ^2 }\over {g_a^2(Q^2) }} \ =\ ReS\
\ +\ {1\over 2}  {{\beta _a}}log{{2M_{string}^6}\over {M_c^4\alpha Q ^2}}
\eeq
This equation is telling us that the one-loop corrected couplings
behave in an {\it effective} manner as if there was
standard field theory logarithmic running not only
up to the scale $M_{string}$ but up to a virtual scale
$M_X$ defined by:
\beq
M_X \ =\ { {\sqrt{2}M_{string}^3} \over {M_c^2\sqrt{\alpha } } }\  =  \
\sqrt{\alpha } M_{Planck}{{M_c}\over
{M_{string} }}\
\label{mx}
\eeq
where we have used the equation $M_{Planck}=(\sqrt{2}/\alpha )M_{string}^4/M_c^3$
(see e.g., ref.\cite{imr} ).
Thus in $Z_N$ orientifolds of the class here discussed, there is
acually "mirage unification" in the sense described in the introduction.

\section{An extension of the MSSM with mirage unification}

The above discussion was made in the context of $Z_N$ compact
Type IIB $N=1$, $D=4$ $Z_N$ orientifolds with odd $N$.
In the case of even $N$ the cancelation off $\sigma $-model
anomalies is expected for some of the three complex planes but
the situation concerning the others is not clear \cite{iru2} .
Furthermore, no completely realistic model has been yet obtained
from this type of Type IIB constructions.

Nevertheless the mechanism found for this class of orientifolds
is quite elegant and something similar to it could be at work
in a more realistic model. Thus, motivated
by the mechanism above,  we would like to present
a simple extension of the MSSM incorporating it .

The ingredients of the model are as follows:

{\bf i)} The gauge group will be $SU(3)\times SU(2)\times U(1)_Y\times U(1)_X$,
where the $U(1)_X$ has to be an anomalous symmetry with anomaly
cancelled by a Green-Schwarz mechanism. We do not need to commit ourselves
for the moment with an specific charge asignement. Notice only that
the mixed anomalies of this $U(1)_X$ with the SM groups are necesarily  unequal,
as is the general  case in orientifolds. They are however very much
constrained, as we discuss below.

{\bf ii)} We assume the Kahler potential presents a classical $SL(2,{\bf R})$
invariance and the charged dilaton $S$, overall modulus $T$ and
MSSM charged chiral fields $\phi _a$ have a Kahler potential of the form
\footnote{Actually the structure of the metric of charged fields could be
different as long as the $\sigma $-model anomalies are proportional to the
$\beta$-functions.}
\beq
K(S,S^*,T,T^*,\phi_{\alpha }, \phi _{\alpha }^*)\ =\ - log(S+S^*)-3\log(T+ T^*)
\ +\ \sum_{\alpha } { {\phi_{\alpha }\phi_{\alpha }^*}\over {T+T^*} }
\label{kahler}
\eeq
where the sum on $\alpha $ goes over all the charged chiral fields of the MSSM.

{\bf iii)}
The $\sigma $-model and $U(1)_X$ anomalies are cancelled by
a Green-Schwarz mechanism involving a singlet $M$ field.
 This singlet field appears in the gauge kinetic function in
the form (in analogy with the orientifold results):
\beq
f_a\ =\ S \ +\ { {\beta _a}\over 2} M
\label{efe}
\eeq
and transforms under $U(1)_X$ and $SL(2,{\bf R})$ transformations as:
\beqa
ImM\ &  \rightarrow &  \ ImM\  + \ \delta_{GS}^X \ \Lambda_X(x) \\
ImM\  & \rightarrow  & \ ImM\ -\ 2log(icT+d)
\label{transon}
\eeqa
where $\delta_{GS}^X$ is a coefficient which would depend on the
$U(1)_X$ charge asignements. Notice that the $\sigma $-model
anomaly coefficients  computed from
eqs.(\ref{coeff}) and (\ref{kahler})  are  given by the $\beta $-functions.
Thus, in this simple model with a single $U(1)$ and a single $M$-field,
the mixed $U(1)_X$-SM standard model anomaly coefficients $A_X^a$
must obey:
\beq
 A_X^a \ =\ -\delta_{GS}^X {{\beta_a}\over 2}
\label{anomon}
\eeq
and  the mixed anomalies must be in the same ratio
as the beta-function coefficients. This was indeed the
case in $Z_N$ , $N$-odd orientifolds, as we showed above.
This is quite a restrictive condition for the charge
asignements of the anomalous $U(1)$ charges.

{\bf iv)}
The singlet $M$ field will have a Kahler potential of the form:
\beq
K(M,M^*) \ =\ (M+M^* \ -\ \delta_{GS}^X V_X \ -\ 2 log(T+T^*))^2
\label{kahlerm}
\eeq
so that it is invariant under both $U(1)$ and $\sigma$-model invariance.

With the above four ingredients results very analogous to the
ones found for the orientifolds are obtained.
The Fayet-Iliopoulos term for the anomalous $U(1)_X$ will be given by:
\beq
\xi_X \  =\  -   \delta_{GS}^X \ (M+M^*
\ - \ 2log(T+T^*)   )
\label{fiagranel2}
\eeq
Since we are interested in studying the corrections for
the gauge couplings of the {\it unbroken} SM group,
we will study   the $\xi _X=0$ field theory direction.
Minimization of the scalar potential will
then  require   $ReM=log(T+T^*)$. Thus we will have
substituting in the real part of  (\ref{efe})
\beq
 {{8\pi^2}\over g_a^2}\  = ReS \ + \ {1\over 2} \beta _a  log(T+T^*)
\label{ffuncdesm}
\eeq
 Now, using the definition of $ReT$ when gauge fields live on 3-branes,
$T+T^*=2M_{string}^4/(M_c^4\alpha )$ one finds after including
the running up to the string scale:
\beq
{{8\pi ^2 }\over {g_a^2(Q^2) }}  \ =\ ReS\
\ +\ {1\over 2}  {{\beta _a}}log{{M_X^2}\over { Q ^2}}
\eeq
with the virtual scale $M_X$ given by:
\beq
M_X \ =\ { {\sqrt{2}M_{string}^3} \over {M_c^2\sqrt{\alpha } } }\  =  \
\sqrt{\alpha } M_{Planck}{{M_c}\over
{M_{string} }}\
\label{mx2}
\eeq
Now,  in this simple extension of the MSSM standard
agreement of gauge coupling unification is achieved
as long as  $M_X=2\times 10^{16}$
GeV. This result may only be obtained for
\footnote{Notice we have also the constraint
 $M_{Planck}=(\sqrt{2}/\alpha )M_{string}^4/M_c^3$ so that
$M_{string}$ and $M_c$ are uniquely fixed in terms of $M_X$ and
$M_{Planck}$.}
\beqa
M_{string}\ & = & {1\over {\sqrt{2\alpha }} }
{{M_X^3}\over {M_{Planck}^2 } }\ =\ 2\times 10^{11} \ GeV \\
M_c\ & = &\ {1\over {\alpha \sqrt{2}} }
{{M_X^4}\over {M_{Planck}^3 } }\ =\ 1.6 \times 10^{9} \ GeV
 \label{msmc}
\eeqa
%
\begin{figure}
\centering
\epsfxsize=3.5in
\hspace*{0in}\vspace*{.2in}
\epsffile{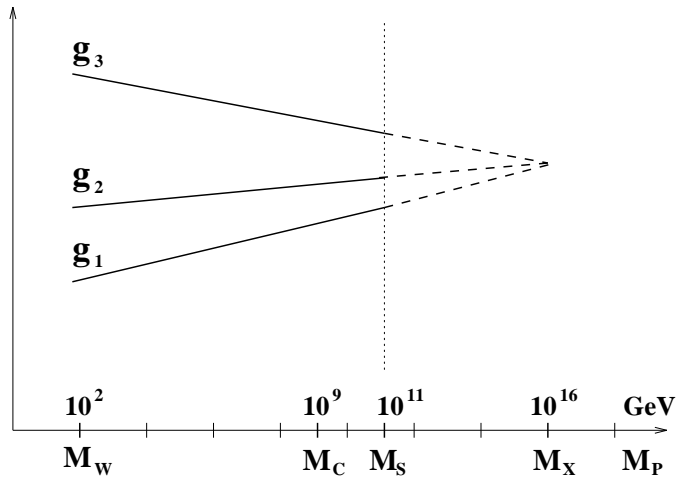}
\caption{\small Mirage unification in the MSSM. The couplings run up to
the string scale $M_s\approx 10^{11} GeV$. The compactification scale
$M_c\approx 10^9 GeV$ creates no new KK thresholds, since the gauge fields
live on 3-branes. The couplings have an apparent unification at the
virtual scale $M_X=\sqrt{\frac {2}{\alpha}} M_s^3/M_c^2$. From low
energies everything looks like if there was a field theory desert in
between $M_W$ and $M_X$.
}
\label{mirage}
\end{figure}
Notice in particular that the ratio
$M_X/M_{Planck}=\sqrt{\alpha}M_c/M_{string}$ so that in the present scheme
the well known missmatch between  Planck mass and 
(virtual) unification scale
$M_X$ would be a reflection of an analogous missmatch between string scale
and compactification scale.

Remarkably enough the above  values for
the fundamental
scales were suggested on the basis of
completely different arguments in ref.\cite{BIQ}  .
Indeed, if one assumes that
SUSY is broken in a far away 3-brane non-SUSY hidden sector and
it is transmitted only by bulk fields to the "observable 3-brane"
world, one expects soft SUSY-breaking terms to be generated
of order $M_{SB}=M_{string}^2/M_{Planck}$
$= \alpha M_c^3/(\sqrt{2} M_{string}^2)$.
These soft terms will trigger $SU(2)\times U(1)$ breaking and
hence their size is of order $M_W$. On the other hand
one can write in general $M_{Planck}=\sqrt{2}M_{string}^4/(\alpha  M_c^3)$.
One thus obtains $M_W/M_{Planck}=\alpha ^2/2(M_c/M_{string})^6$ in
this situation. Now, if $M_c/M_{string}\propto 1/160$ one can understand
the huge $M_W/M_{Planck}$ hierarchy in terms of the modest ratio
$M_c/M_{string}$. It is quite satisfactory to find that "mirage
gauge coupling unification" naturally requires the same distribution
of mas scales, much more so since this was not our initial motivation.

 Notice that in an isotropical  1 TeV string scenario
\cite{untev,anton}
chosing $M_c=10^{-2}$GeV and $M_{string}=10^3$ GeV gives $M_X=10^{13}$
GeV, which would be too low.

Due to the form of (\ref{kahlerm}) one can check that the $U(1)_X$ gauge
boson gets a mass of order the string scale $M_{string}$ \cite{pop}.
 The same happens
with a linear combination of $M$ and $T$. The orthogonal linear combination
remains massless at this level and, in particular, its imaginary part
will have axion-like couplings and might help
\cite{BIQ} to solve the strong CP problem.

\section{Comments and outlook}

We have described above how in a  class
of Type IIB , $D=4$, $N=1$ orientifolds a peculiar phenomenon occurrs
concerning gauge coupling unification: although gauge coupling running
occurs only up to the string scale $M_{string}$, there
are  modulus-dependent corrections which mimic further running up to
a virtual scale $M_X=\sqrt{\alpha }M_{Planck}M_c/M_{string}$.
The  modulus dependence appears due to the simultaneous presence of
anomalous $U(1)$'s and $\sigma $-model symmetries in this class of theories.

It would be interesting to know how general this property is.
The status  of $\sigma $-model symmetries in other
classes of orientifolds like  e.g. those with $N$ even is less clear
so that a direct application of our argumentation does  not necesarily work.
However it could well be that the final result is quite independent of
the derivation. Certainly, the appearence of large $T$-dependent
corrections in the gauge coupling constants seems generic.
In  compact orientifolds it seems generic
the presence of $T$-moduli dependent pieces in the Fayet-Iliopoulos
terms $\xi_a$. Upon minimization  one could  be able to express
$ReM=ReM(T,T^*)$ which when substituted back into the gauge kinetic function
(whose $M$-dependence is also generic) will give rise
in principle  to large $T$-dependent corrections to gauge coupling constants.
The least one has to say is that
{\it  these effects cannot in general be neglected } and have to be taken into
account before giving any account of gauge coupling unification.
This, however does not imply that  these
$T$-dependent corrections will
get precisely the $\beta $-function coefficient required to get
"mirage unification".  The mechanism nevertheless looks quite general
and very likely will be present in other classes of vacua different from the
$N$ odd orientifolds. Notice in this connection that in order to obtain
"mirage unification" it is enough that the {\it difference} between the
$M$-dependent corrections to the gauge coupling constants have  a
coefficient proportional to the {\it difference } between the
respective $\beta $-functions.

>From the phenomenological point of view, "mirage unification" offers an
elegant and possibly unique option to lower the string scale and still
mantaining the success of gauge coupling unification to the {\it
same level of agreement and predictivity to that of  SUSY-GUT's} .
In the previous section we showed  how a simple  modification of the
MSSM to include some "closed string" fields $S$, $T$ and $M$ and
simultaneous presence of an anomalous $U(1)$ and  $\sigma $-model
invariance can give rise to the required mechanism.
Thus the mechanism itself is quite general.  The anomalous
$U(1)$ is very much restricted in the simplest model
(only one $U(1)$ and only one $M$ field) since its mixed
anomalies with $SU(3)\times SU(2)\times U(1)_Y$
have to be in the ratio of the corresponding $\beta $ functions.
This may lead to quite interesting constraints for model building
which are quite distinct to those worked out for heterotic models
\cite{sin,ramond}
in which those anomalies are in the ratio of the coupling constant
normalizations.
Notice also that both the presence of the $U(1)$ and the $\sigma $-model symmetry
are wellcome in order to supress sufficiently proton decay
mediated by dimension four or higher operators.

We have found that if one wants to identify the "virtual unification scale"
$M_X=\sqrt{\alpha }M_{Planck}M_c/M_{string}$ with the scale suggested by
experiment,  one necesarily has to use  as inputs
$M_{string}=10^{11}$ GeV and $M_c=10^{9}$. Thus in mirage gauge coupling
unification the coupling constants give us a measure of the fundamental scales
of the theory, $M_{string}$ and $M_c$.
This is an unexpected
result of the present analysis which fits quite well with the suggestion of
ref.\cite{BIQ}  to identify $M_{string}$ with the intermediate scale
$\sqrt{M_{Planck}M_W}=10^{11}$ GeV.  We find this fact very intriguing,
particularly so since it was not our intention to find such a connection.

Gauge coupling unification within the MSSM has been always thought to be
a great success and a strong indication of the existence of a unification scale
of order $M_X \propto 10^{16}$ GeV. If the mechanism we suggest is at work,
nature has been a bit nasty with us giving us a (partially) wrong track
pointing towards a big desert in between $M_W$ and $M_X$. We would have been
too naive in assuming that all logarithms come from field-theory running.

\bigskip

\bigskip

\bigskip

\newpage

\bigskip

\centerline{\bf Acknowledgements}
We are grateful to A.~Uranga ,  R. Rabadan  and specially F.Quevedo
for discussions.
L.E.I.  thanks  CICYT (Spain) and
the European Commission (grant ERBFMRX-CT96-0045)
for financial support.

\end{document}